\documentstyle[amsfonts,amssymb,aps,twocolumn]{revtex}

\input{epsf}

\title{Achievable rates for the Gaussian quantum channel\thanks{CALT-68-2323}}
\author{Jim Harrington\thanks{\tt
jimh@theory.caltech.edu} and John Preskill\thanks{\tt
preskill@theory.caltech.edu}}

\address{Institute for Quantum Information, California Institute of Technology, 
Pasadena, CA 91125, USA}

\begin{document}

\maketitle

\begin{abstract}
We study the properties of quantum stabilizer codes that embed a finite-dimensional protected code space in an infinite-dimensional Hilbert space. The stabilizer group of such a code is associated with a symplectically integral lattice in the phase space of $2N$ canonical variables. From the existence of symplectically integral lattices with suitable properties, we infer a lower bound on the quantum capacity of the Gaussian quantum channel that matches the one-shot coherent information optimized over Gaussian input states.
 
\end{abstract}

\parskip=5pt
\section{Introduction}

A central problem in quantum information theory is to determine the quantum capacity of a noisy quantum channel --- the maximum rate at which coherent quantum information can be transmitted through the channel and recovered with arbitrarily good fidelity \cite{nielsen_chuang,preskill229}. A general solution to the corresponding problem for classical noisy channels was found by Shannon in the pioneering paper that launched classical information theory \cite{shannon,cover}. With the development of the theory of quantum error correction \cite{shor_9,steane_7}, considerable progress has been made toward characterizing the quantum channel capacity \cite{ibm}, but it remains less well understood than the classical capacity. 

The asymptotic coherent information has been shown to provide an upper bound on the capacity \cite{schumacher,barnum} and a matching lower bound has been conjectured, but not proven \cite{lloyd}. Unfortunately, the coherent information is not subadditive \cite{shor_smolin}, so that its asymptotic value is not easily computed. Therefore, it has been possible to verify the coherent information conjecture in just a few simple cases \cite{ibm_erasure}.  

One quantum channel of considerable intrinsic interest is the Gaussian quantum channel, which might also be simple enough to be analytically tractable, thus providing a fertile testing ground for the general theory of quantum capacities.  A simple analytic formula for the capacity of the Gaussian classical channel was found by Shannon \cite{shannon,cover}. The Gaussian quantum channel was studied by Holevo and Werner \cite{holevo}, who computed the one-shot coherent information for Gaussian input states, and derived an upper bound on the quantum capacity. 

Lower bounds on the quantum capacity of the Gaussian quantum channel were established by Gottesman, Kitaev and Preskill \cite{gott_kit_pres}. They developed quantum error-correcting codes that protect a finite-dimensional subspace of an infinite-dimensional Hilbert space, and showed that these codes can be used to transmit high-fidelity quantum information at a nonzero asymptotic rate. In this paper, we continue the study of the Gaussian quantum channel begun in \cite{gott_kit_pres}. Our main result is that the coherent information computed by Holevo and Werner is in fact an achievable rate. This result lends nontrivial support to the coherent information conjecture. 

We define the Gaussian quantum channel and review the results of Holevo and Werner \cite{holevo} in Sec. II. In Sec. III, we describe the stabilizer codes for continuous quantum variables introduced in \cite{gott_kit_pres}, which are based on the concept of a symplectically integral lattice embedded in phase space. In Sec. IV and V, we apply these codes to the Gaussian quantum channel, and calculate an achievable rate arising from lattices that realize efficient packings of spheres in high dimensions. This achievable rate matches the one-shot coherent information $I_Q$ of the channel in cases where $2^{I_Q}$ is an integer. Rates achieved with concatenated coding are calculated in Sec. VI; these fall short of the coherent information but come close. In Sec. VII, we consider the Gaussian classical channel, and again find that concatenated codes achieve rates close to the capacity. Sec. VIII contains some concluding comments about the quantum capacity of the Gaussian quantum channel.

\section{The Gaussian quantum channel}

The Gaussian quantum channel is a natural generalization of the Gaussian classical channel.
In the classical case, we consider a channel such that the input $x$ and the output $y$ are real numbers. The channel applies a displacement to the input by distance $\xi$,
\begin{equation}
y=x+\xi~,
\end{equation}
where $\xi$ is a Gaussian random variable with mean zero and variance $\sigma^2$; the probability distribution governing $\xi$ is 
\begin{equation}
P(\xi)= {1\over\sqrt{2\pi\sigma^2}}e^{-\xi^2/2\sigma^2}~.
\end{equation}

Similarly, acting on a quantum system described by canonical variables $q$ and $p$ that satisfy  the commutation relation $[q,p]=i\hbar$, we may consider a quantum channel that applies a phase-space displacement described by the unitary operator
\begin{equation}
D(\alpha)= \exp\left( \alpha a^\dagger + \alpha^* a\right)~,
\end{equation}
where $\alpha$ is a complex number, $[a,a^\dagger]=1$, and $q$, $p$ can be expressed in terms of $a$ and $a^\dagger$ as
\begin{equation}
q= \sqrt{\hbar\over 2}\left(a+a^\dagger\right)~,\quad p= -i\sqrt{\hbar\over 2}\left(a-a^\dagger\right)~.
\end{equation}
This quantum channel is Gaussian if $\alpha$ is a complex Gaussian random variable with mean zero and variance $\tilde\sigma^2$. In that case, the channel is the superoperator (trace-preserving completely positive map) ${\cal E}$ that acts on the density operator $\rho$ according to
\begin{equation}
\rho\to {\cal E}(\rho)= {1\over \pi\tilde \sigma^2}\int d^2\alpha ~e^{-|\alpha|^2/\tilde\sigma^2}D(\alpha)\rho D(\alpha)^\dagger ~.
\end{equation}
In other words, the position $q$ and momentum $p$ are displaced independently,
\begin{equation}
q\to q +\xi_q~,\quad p\to p +\xi_p~,
\end{equation}
where $\xi_q$ and $\xi_p$ are real Gaussian random variables with mean zero and variance $\sigma^2=\hbar\tilde\sigma^2$.

To define the capacity, we consider a channel's $n$th extension. In the classical case, a message is transmitted consisting of the $n$ real variables
\begin{equation}
\vec x= (x_1,x_2, \dots, x_n)~,
\end{equation}
and the channel applies the displacement
\begin{equation}
\vec x \to \vec x + \vec \xi ~, \quad \vec \xi=(\xi_1,\xi_2,\dots, \xi_n)~,
\end{equation}
where the $\xi_i$'s are independent Gaussian random variables, each with mean zero and variance $\sigma^2$. A code consists of a finite number $m$ of $n$-component input signals
\begin{equation}
\vec x^{(a)}~, \quad a=1,2, \dots, m
\end{equation}
and a decoding function that maps output vectors to the index set $\{1,2,\dots, m\}$. We refer to $n$ as the {\em length} of the code.

If the input vectors were unrestricted, then for fixed $\sigma^2$ we could easily construct a code with an arbitrarily large number of signals $m$ and a decoding function that correctly identifies the index $(a)$ of the input with arbitrarily small probability of error; even for $n=1$ we merely choose the distance between signals to be large compared to $\sigma$. To obtain an interesting notion of capacity, we impose a constraint on the {\em average power} of the signal,
\begin{equation}
{1\over n}\sum_i \left(x_i^{(a)}\right)^2 \le P~,
\end{equation} 
for each $a$. We say that a rate $R$ (in bits) is achievable with power constraint $P$ if the there is a sequence of codes satisfying the constraint such that the $\beta$th code in the sequence contains $m_\beta$ signals with length $n_\beta$, where
\begin{equation}
R=\lim_{\beta\to\infty}~  {1\over n_\beta} \log_2 m_\beta~,
\end{equation}
and the probability of a decoding error vanishes in the limit $\beta\to \infty$. The capacity of the channel with power constraint $P$ is the supremum of all achievable rates.

The need for a constraint on the signal power to define the capacity of the Gaussian classical channel can be understood on dimensional grounds. The classical capacity (in bits) is a dimensionless function of the variance $\sigma^2$, but $\sigma^2$ has dimensions. Another quantity with the dimensions of $\sigma^2$ is needed to construct a dimensionless variable, and the power $P$ fills this role.

In contrast, no power constraint is needed to define the quantum capacity of the quantum channel. Rather, Planck's constant $\hbar$ enables us to define a dimensionless variance $\tilde\sigma^2=\sigma^2/\hbar$, and the capacity is a function of this quantity. In the quantum case, a code consists of an encoding superoperator that maps an $m$-dimensional Hilbert space ${\cal H}_m$ into the infinite-dimensional Hilbert space ${\cal H}^{\otimes N}$ of $N$ canonical quantum systems, and a decoding superoperator that maps ${\cal H}^{\otimes N}$ back to ${\cal H}_m$. We say that the rate $R$ (in qubits) is achievable if there is a sequence of codes such that
\begin{equation}
R=\lim_{\beta\to\infty}~  {1\over N_\beta} \log_2 m_\beta~,
\end{equation}
where arbitrary states in ${\cal H}_m$ can be recovered with a fidelity that approaches 1 as $\beta\to\infty$. The quantum capacity $C_Q$ of the channel is defined as the supremum of all achievable rates.

Holevo and Werner \cite{holevo} studied a more general Gaussian channel that includes damping or amplification as well as displacement. However, we will confine our attention in this paper to channels that apply only displacements. Holevo and Werner derived a general upper bound on the quantum capacity by exploiting the properties of the ``diamond norm'' (norm of complete boundedness) of a superoperator. The diamond norm is defined as follows: First we define the trace norm of an operator $X$ as 
\begin{equation}
\| X \|_{\rm tr}\equiv {\rm tr}\sqrt{X^\dagger X}~,
\end{equation}
which for a self-adjoint operator is just the sum of the absolute values of the eigenvalues. Then a norm of a superoperator ${\cal E}$ can be defined as 
\begin{equation}
\| {\cal E}\|_{\rm so}= \sup_{X\ne0} {\| {\cal E}(X)\|_{\rm tr}\over \| X\|_{\rm tr}}~.
\end{equation}
The superoperator norm is not stable with respect to appending an ancillary system on which ${\cal E}$ acts trivially. Thus we define the diamond norm of ${\cal E}$ as
\begin{equation}
\| {\cal E}\|_{\diamond}=\sup_n \| {\cal E}\otimes I_n\|_{\rm so}~,
\end{equation}
where $I_n$ denotes the $n$-dimensional identity operator. (This supremum is always attained for some $n$ no larger than the dimension of the Hilbert space on which ${\cal E}$ acts.) Holevo and Werner showed that the quantum capacity obeys the upper bound
\begin{equation}
C_Q({\cal E})\le \log_2 \| {\cal E}\circ T\|_\diamond~,
\end{equation}
where $T$ is the transpose operation defined with respect to some basis. In the case of the Gaussian quantum channel, they evaluated this expression, obtaining
\begin{equation}
C_Q(\sigma^2)\le \log_2\left(\hbar/\sigma^2\right)~
\end{equation}
for $\hbar/\sigma^2 > 1$, and $C_Q(\sigma^2)=0$ for $\hbar/\sigma^2\le 1$.

Holevo and Werner \cite{holevo} also computed the {\em coherent information} of the Gaussian quantum channel for a Gaussian input state. To define the coherent information of the channel ${\cal E}$ with input density operator $\rho$, one introduces a reference system $R$ and a {\em purification} of $\rho$, a pure state $|\Phi\rangle$ such that
\begin{equation}
{\rm tr}_R\left(|\Phi\rangle\langle \Phi|\right)=\rho~.
\end{equation}
Then the coherent information $I_Q$ is
\begin{equation}
I_Q({\cal E},\rho)= S\left({\cal E}(\rho)\right) - S\left({\cal E}\otimes I_R(|\Phi\rangle\langle \Phi|)\right)~,
\end{equation}
where $S$ denotes the Von Neumann entropy,
\begin{equation}
S(\rho)=-{\rm tr}\left(\rho \log_2 \rho\right)~.
\end{equation}
It is {\em conjectured} \cite{lloyd,schumacher,barnum} that the quantum capacity is related to the coherent information by
\begin{equation}
C_Q({\cal E})= \lim_{n\to\infty}~{1\over n}\cdot C_n({\cal E})~,
\end{equation}
where
\begin{equation}
C_n({\cal E})= \sup_{\rho} I_Q({\cal E}^{\otimes n},\rho)~.
\end{equation}
Unlike the mutual information that defines the classical capacity, the coherent information is not subadditive in general, and therefore the quantum capacity need not coincide with the ``one-shot'' capacity $C_1$. 
Holevo and Werner showed that for the Gaussian quantum channel, the supremum of $I_Q$ over Gaussian input states is
\begin{equation}
\label{hw_coherent}
\left(I_Q\right)_{\rm max}= \log_2(\hbar/e\sigma^2)~
\end{equation}
(where $e=2.71828..$) for $\hbar/e\sigma^2 > 1$, and $\left(I_Q\right)_{\rm max} =0$ for $\hbar/e\sigma^2 \le 1$. According to the coherent-information conjecture, eq.~(\ref{hw_coherent}) should be an achievable rate.

\section{Quantum error correcting codes for continuous quantum variables}
The lattice codes developed in \cite{gott_kit_pres} are stabilizer codes \cite{gottesman,att} that embed a finite-dimensional code space in the infinite-dimensional Hilbert space of $N$ ``oscillators,'' a system described by $2N$ canonical variables $q_1,q_2,\dots q_N,p_1,p_2,\dots,p_N$. That is, the code space is the simultaneous eigenstate of $2N$ commuting unitary operators, the generators of the code's stabilizer group. Each stabilizer generator is a {\em Weyl operator}, a displacement in the $2N$-dimensional phase space. 

Such displacements can be parametrized by $2N$ real numbers $\alpha_1,\alpha_2,\dots, \alpha_N,\beta_1,\beta_2,\dots,\beta_N$, and expressed as
\begin{equation}
U(\alpha,\beta)=\exp\left[i\sqrt{2\pi}\left(\sum_{i=1}^N \alpha_i p_i + \beta_i q_i\right)\right]~.
\end{equation}
Two such operators obey the commutation relation
\begin{equation}
\label{weyl_com}
U(\alpha,\beta)U(\alpha',\beta')=e^{2\pi i\omega(\alpha\beta,\alpha'\beta')} U(\alpha',\beta')U(\alpha,\beta)~,
\end{equation}
where 
\begin{equation}
\omega(\alpha\beta,\alpha'\beta')\equiv \alpha\cdot\beta'-\alpha'\cdot\beta
\end{equation}
is the symplectic form. Thus Weyl operators commute if and only if their symplectic form is an integer.

The $2N$ generators of a stabilizer code are commuting Weyl operators
\begin{equation}
U\left(\alpha^{(a)},\beta^{(a)}\right)~,\quad a=1,2,\dots, 2N~.
\end{equation}
Thus the elements of the stabilizer group are in one-to-one correspondence with the points of a lattice ${\cal L}$ generated by the $2N$ vectors $v^{(a)}=(\alpha^{(a)},\beta^{(a)})$. These vectors can be assembled into the generator matrix $M$ of ${\cal L}$ given by
\begin{equation} 
M=\pmatrix{v^{(1)}\cr
v^{(2)}\cr\cdot\cr\cdot\cr
v^{(2N)}}~.
\end{equation} 
Then the requirement that the stabilizer generators commute, through eq.~(\ref{weyl_com}), becomes the condition that the antisymmetric matrix
\begin{equation}
A= M\omega M^T
\end{equation}
has integral entries, where $M^T$ denotes the transpose of $M$, $\omega$ is the $2N\times 2N$ matrix
\begin{equation}
\omega=\pmatrix{0& I_N\cr -I_N& 0}
\end{equation}
and $I_N$ is the $N\times N$ identity matrix. If the generator matrix $M$ of a lattice ${\cal L}$ has the property that $A$ is an integral matrix, then we will say that the lattice ${\cal L}$ is {\em symplectically integral}.

Encoded operations that preserve the code subspace are associated with the code's {\em normalizer} group, the group of phase space translations that commute with the code stabilizer. The generator matrix of the normalizer is a matrix $M^\perp$ that can be chosen to be
\begin{equation}
M^\perp = A^{-1}M~,
\end{equation}
so that
\begin{equation}
M^\perp \omega M^T = I~;
\end{equation}
and 
\begin{equation}
\left(M^\perp\right) \omega \left(M^\perp\right)^T= \left(A^{-1}\right)^T~.
\end{equation}
We will refer to the lattice generated by $M^\perp$ as the {\em symplectic dual} ${\cal L}^\perp$ of the lattice ${\cal L}$.

Another matrix that generates the same lattice as $M$ (and therefore defines a different set of generators for the same stabilizer group) is
\begin{equation}
M'= RM~,
\end{equation}
where $R$ is an integral matrix with $\det R=\pm 1$. This replacement changes the matrix $A$ according to
\begin{equation}
A\to  RAR^T~.
\end{equation}
By Gaussian elimination, an $R$ can be constructed such that 
\begin{equation}
A=\pmatrix{0& D\cr -D& 0}~,
\end{equation}
and
\begin{equation}
\left(A^{-1}\right)^T= \pmatrix{0& D^{-1}\cr -D^{-1} &0}~,
\end{equation}
where $D$ is a positive diagonal integral $N\times N$ matrix. In the important special case of a {\em symplectically self-dual} lattice, both $A$ and $\left(A^{-1}\right)^T$ are integral matrices; therefore $D=D^{-1}$ and the standard form of $A$ is
\begin{equation}
A=\pmatrix{0& I_N\cr -I_N& 0}=\omega~.
\end{equation}
Hence the generator matrix of a symplectically self-dual lattice can be chosen to be a real symplectic matrix: $M\omega M^T=\omega$.

If the lattice is rotated, then the generator matrix is transformed as
\begin{equation}
M\to MO~,
\end{equation}
where $O$ is an orthogonal matrix. Therefore, it is convenient to characterize a lattice with its Gram matrix
\begin{equation}
G=MM^T~,
\end{equation}
which is symmetric, positive, and rotationally invariant. In the case of a symplectically self-dual lattice, the Gram matrix $G$ can be chosen to be symplectic, and 
two symplectic Gram  matrices $G$ and $G'$ describe the same lattice if
\begin{equation}
G'= RGR^T~,
\end{equation}
where $R$ is symplectic and integral. Therefore, the moduli space of symplectically self-dual lattices in $2N$ dimensions can be represented as 
\begin{equation}
{\cal A}_N=H(2N)/Sp(2N,Z)~,
\end{equation}
where $H(2N)$ denotes the space of real symplectic positive $2N\times 2N$ matrices of determinant 1. The space ${\cal A}_N$ can also be identified as the moduli space of  principally polarized abelian varieties in complex dimension $N$ \cite{sarnak}. 

The encoded operations that preserve the code space but act trivially within the code space comprise the quotient group ${\cal L}^\perp/{\cal L}$. The order of this group, the ratio of the volume of the unit cell of ${\cal L}$ to that of ${\cal L}^\perp$, is $m^2$, where $m$ is the dimension of the code space. The volume of the unit cell of ${\cal L}$ is $|\det M|= |\det A|^{1/2}$ and the volume of the unit cell of ${\cal L}^\perp$ is $|\det M^\perp|= |\det A|^{-1/2}$; therefore
the dimension of the code space is
\begin{equation}
m= |{\rm Pf}~ A| = |\det M|= \det D~,
\end{equation}
where ${\rm Pf}~ A$ denotes the Pfaffian of $A$, the square root of its determinant. Thus, a symplectically self-dual lattice, for which $|\det M|= |\det M^\perp|=1$, corresponds to a code with a one-dimensional code space. Given a $2N\times 2N$ generator matrix $M$ of a symplectically self-dual lattice, we can rescale it as
\begin{equation}
M\to \sqrt{\lambda} M~,
\end{equation}
where $\lambda$ is an integer, to obtain the 
generator matrix of a symplectically integral lattice corresponding to a code of dimension
\begin{equation}
m= \lambda^N~.
\end{equation}
The rate of this code, then, is 
\begin{equation}
R= \log_2\lambda~.
\end{equation}

When an encoded state is subjected to the Gaussian quantum channel, a phase space displacement \begin{equation}
(\vec q,\vec p)\to (\vec q,\vec p)+(\vec\xi_q,\vec\xi_p)
\end{equation}
is applied. To diagnose and correct this error, the eigenvalues of all stabilizer generators are measured, which determines the value of $(\vec\xi_q,\vec\xi_p)$ modulo the normalizer lattice ${\cal L}^\perp$. To recover, a displacement of minimal length is applied that returns the stabilizer eigenvalues to their standard values, and so restores the quantum state to the code space. We can associate with the origin of the normalizer lattice its {\em Voronoi cell}, the set of points in ${\Bbb R}^{2N}$ that are closer to the origin than to any other lattice site. Recovery is successful if the applied displacement lies in this Voronoi cell. Thus, we can estimate the likelihood of a decoding error by calculating the probability that the displacement lies outside the Voronoi cell.

\section{Achievable rates from efficient sphere packings}
One way to establish an achievable rate for the Gaussian quantum channel is to choose a normalizer lattice ${\cal L}^\perp$ whose shortest nonzero vector is sufficiently large. In this Section, we calculate an achievable rate by demanding that the  Voronoi cell surrounding the origin contain all typical displacements of the origin in the limit of large $N$. In Sec. V, we will use a more clever argument to improve our estimate of the rate.

The volume of a sphere with unit radius in $n$ dimensions is
\begin{equation}
V_n= {{\pi^{n/2}}\over \Gamma\left({n\over 2}+1\right)}~,
\end{equation}
and from the Stirling approximation we find that
\begin{equation}
V_n \le \left(2\pi e\over n\right)^{n/2}~. 
\end{equation}
It was shown by Minkowski \cite{minkowski} that lattice sphere packings exist in $n$ dimensions that fill a fraction at least $1/2^{(n-1)}$ of space. Correspondingly, if the lattice is chosen to be unimodular, so that its unit cell has unit volume, then kissing spheres centered at the lattice sites can be chosen to have a radius $r_n$ such that
\begin{equation}
V_n \left(r_n\right)^n \ge 2^{-(n-1)}~,
\end{equation}
or 
\begin{equation}
r_n^2 \ge {1\over 4} (2/V_n)^{2/n} \ge {n\over 8\pi e}~.
\end{equation}
This lower bound on the efficiency of sphere packings has never been improved in the nearly 100 years since Minkowski's result. More recently, Buser and Sarnak \cite{sarnak} have shown that this same lower bound applies to lattices that are symplectically self-dual.

Now consider the case of $n=2N$-dimensional phase space. For sufficiently large $n$, the channel will apply a phase space translation by a distance which with high probability will be less than $\sqrt{n(\sigma^2+\varepsilon)}$, for any positive $\varepsilon$. Therefore, a code that can correct a shift this large will correct all likely errors. What rate can such a code attain? If the code is a lattice stabilizer code, and the dimension of the code space is $m$, then the unit cell of the code's normalizer lattice has volume
\begin{equation}
\Delta= {1\over m}\cdot (2\pi\hbar)^N~.
\end{equation}
Nonoverlapping spheres centered at the sites of the normalizer lattice can be chosen to have radius $r=\sqrt{n(\sigma^2 +\varepsilon)}$, where
\begin{equation}
\left({2\pi e\over n}\right)^{n/2} \left(n(\sigma^2 + \varepsilon)\right)^{n/2} \ge {1\over m}\cdot 2^{-n}\cdot ({2\pi\hbar})^{n/2}~,
\end{equation}
or
\begin{equation}
\label{dimen_epsilon}
m\ge \left({\hbar \over 4e(\sigma^2 +\varepsilon)}\right)^N~.
\end{equation}
The error probability becomes arbitrarily small for large $N$ if eq.~(\ref{dimen_epsilon}) is satisfied, for any positive $\varepsilon$. We conclude that the rate
\begin{equation}
\label{nonoverlap_rate}
R\equiv {1\over N} \cdot \log_2 m = \log_2\left({\hbar\over 4e\sigma^2}\right)~,
\end{equation}
is achievable, provided $\hbar/4e\sigma^2 \ge 1$.
However, as noted in Sec. III, the rates that can be attained by this construction (rescaling of a symplectically self-dual lattice) are always of the form $\log_2\lambda$, where $\lambda$ is an integer.

\section{Improving the rate}
\label{sec:improve}
\begin{figure}
\begin{center}
\leavevmode
\epsfxsize=3in
\epsfbox{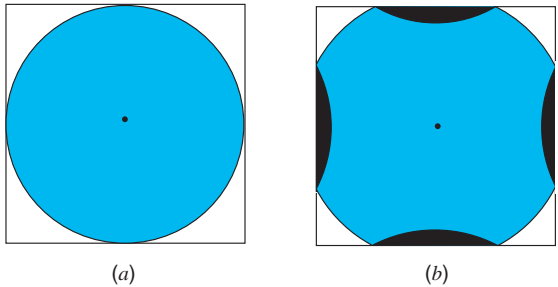}
\end{center}
\caption{Two ways to estimate the rate achieved by a lattice code. Each site of the normalizer lattice has a Voronoi cell (represented here by a square) containing all points that are closer to that site than any other site. Displacements that move a site to a position within its Voronoi cell can be corrected. The volume of the Voronoi cell determines the rate of the code. In $(a)$, the ball containing typical displacements lies within the cell, so that the error probability is small. In $(b)$, the ball of typical displacements is not completely contained within the cell, but the region where neighboring balls overlap (shown in black) has a small volume, so that the error probability is still small.}
\label{fig:square_circle}
\end{figure}

The achievable rate found in eq.~(\ref{nonoverlap_rate}) falls two qubits short of the coherent information eq.~(\ref{hw_coherent}). We will now show that this gap can be closed by using tighter estimates of the error probability. We established eq.~(\ref{nonoverlap_rate}) by filling phase space with nonoverlapping spheres, which is overly conservative. It is acceptable for the spheres to overlap, as long as the overlaps occupy an asymptotically negligible fraction of the total volume, as suggested in Fig.~\ref{fig:square_circle}. 

Our improved estimate applies another result obtained by Buser and Sarnak \cite{sarnak}. They note that the moduli space of symplectically self-dual lattices is compact and equipped with a natural invariant measure. Therefore, it makes sense to consider averaging over all lattices. Denote by $\langle \cdot \rangle$ the average over all symplectically self-dual lattices with specified dimension $n=2N$, and let $f(x)$ denote an integrable rotationally-invariant function of the vector $x$ (that is a function of the length $|x|$ of $x$). Then Buser and Sarnak \cite{sarnak} show that 
\begin{equation}
\Big\langle\sum_{x\in{\cal L }\backslash \{0\}} f(x)\Big\rangle = \int f(x)~ d^{n}x~.
\end{equation}
(Note that the sum is over all {\em nonzero} vectors in the lattice ${\cal L}$.) It follows that there must exist a {\em particular} symplectically self-dual lattice ${\cal L}$ such that
\begin{equation}
\label{mink_hla}
\sum_{x\in{\cal L } \backslash \{0\}} f(x) \le  \int f(x)~ d^{n}x~.
\end{equation}
The statement that a {\em unimodular} lattice exists that satisfies eq.~(\ref{mink_hla}) is the well-known Minkowski-Hlawka theorem \cite{geometry}. Buser and Sarnak established the stronger result that the lattice can be chosen to be symplectically self-dual.
  
We can use this result to bound the probability of a decoding error, and establish that a specified rate is achievable.  Our argument will closely follow de Buda \cite{debuda}, who performed a similar analysis of lattice codes for the Gaussian classical channel. However, the quantum case is considerably easier to analyze, because we can avoid complications arising from the power constraint \cite{shell1,shell2,shell3}.

A decoding error occurs if the channel displaces the origin to a point outside the Voronoi cell centered at the origin. The Voronoi cell has a complicated geometry, so that the error probability is not easy to analyze. But we can simplify the analysis with a trick \cite{debuda}. Imagine drawing a sphere with radius 
\begin{equation}
a=\sqrt{n(\sigma^2 +\varepsilon})
\end{equation}
around each lattice site, where $\varepsilon >0$; this value of $a$ is chosen so that the typical displacement introduced by the channel has length less than  $a$; the probability of a shift larger than $a$ thus becomes negligible for large $n$.  It may be that these spheres overlap. However, a vector that is contained in the sphere centered at the origin, and is not contained in the sphere centered at any other lattice site, must be closer to the origin than any other lattice site. Therefore, the vector is contained in the origin's Voronoi cell, and is a shift that can be corrected successfully. (See Fig.~1.)

Hence (ignoring the possibility of an atypical shift by $\xi > a$) we can upper bound the probability of error by estimating the probability that the shift moves any other lattice site into the sphere of radius $a$ around the origin. We then find
\begin{equation}
P_{\rm error} \le \sum_{x\in{\cal L }^\perp \backslash \{0\}} \int_{|r|\le a}P(x-r)d^n r~,
\end{equation}
where  
$P(\xi)$ denotes the probability of a displacement by $\xi$. 

The Buser-Sarnak theorem \cite{sarnak} tells us that there exists a lattice whose unit cell has volume $\Delta$, and which is related by rescaling to a symplectically self-dual lattice, such that
\begin{equation}
P_{\rm error} \le {1\over\Delta}\int d^n x  \int_{|r|\le a}P(x-r)d^n r~;
\end{equation}
by interchanging the order of integration, we find that 
\begin{equation}
P_{\rm error} \le {1\over\Delta}\cdot V_n\cdot a^n~,
\end{equation}
the ratio of the volume of the $n$-dimensional sphere of radius $a$ to the volume of the unit cell. 

Now the volume $\Delta$ of the unit cell of the normalizer lattice ${\cal L}^\perp$, and the dimension $m$ of the code space, are related by 
\begin{equation}
\Delta=(2\pi\hbar)^N m^{-1}= \left(2\pi\hbar \cdot 2^{-R}\right)^N~,
\end{equation}
where $R$ is the rate, and we may estimate the volume of the sphere as
\begin{equation}
V_n\cdot a^n \le\left({2\pi e\over n}\right)^{n/2}\left(n(\sigma^2 + \varepsilon)\right)^{n/2}~,
\end{equation}
where $n=2N$. Thus we conclude that
\begin{equation}
P_{\rm error} \le \left({e(\sigma^2+\varepsilon)\over \hbar}\cdot 2^R\right)^N~.
\end{equation}
Therefore, the error probability becomes small for large $N$ for any rate $R$ such that
\begin{equation}
R < \log_2\left({\hbar\over e(\sigma^2 +\varepsilon)}\right)~,
\end{equation}
where $\varepsilon$ may be arbitrarily small. We conclude that the rate
\begin{equation}
\label{rate_overlapping}
R = \log_2\left({\hbar\over e\sigma^2}\right)
\end{equation}
is achievable in the limit $N\to \infty$, provided that $\hbar/e\sigma^2>1$. This rate matches the optimal value eq.~(\ref{hw_coherent}) of the one-shot coherent information for Gaussian inputs. We note, again, that the rates that we can obtain from rescaling a symplectically self-dual lattice are restricted to $R=\log_2\lambda$, where $\lambda$ is an integer. Thus for specified $\sigma^2$, the achievable rate that we have established is really the maximal value of
\begin{equation}
R=\log_2\lambda~, \quad \lambda\in Z~,
\end{equation}
such that the positive integer $\lambda$ satisfies
\begin{equation}
\lambda < {\hbar\over e\sigma^2}~.
\end{equation}

\section{Achievable rates from concatenated codes}
\label{sec:concat}

Another method for establishing achievable rates over the Gaussian quantum channel was described in \cite{gott_kit_pres}, based on {\em concatenated coding}. In each of $N$ ``oscillators'' described by canonical variables $p_i$ and $q_i$, a $d$-dimensional system (``qudit'') is encoded that is protected against sufficiently small shifts in $p_i$ and $q_i$. The encoded qudit is associated with a square lattice in 2-dimensional phase space. Then a stabilizer code is constructed that embeds a $k$-qudit code space in the Hilbert space of $N$ qudits; these $k$ encoded qudits are protected if a sufficiently small fraction of the $N$ qudits are damaged. Let us compare the rates achieved by concatenated codes to the rates achieved with codes derived from efficient sphere packings.

We analyze the effectiveness of concatenated codes in two stages. First we consider how likely each of the $N$ qudits is to sustain damage if the underlying oscillator is subjected to the Gaussian quantum channel. The area of the unit cell of the two-dimensional square normalizer lattice that represents the encoded operations acting on the qudit is $2\pi\hbar/d$, and the minimum distance between lattice sites is $\delta=\sqrt{2\pi\hbar/d}$. A displacement of $q$ by $a\cdot \delta$, where $a$ is an integer, is the operation $X^a$ acting on the code space, and a displacement of $p$ by $b\cdot \delta$ is the operation $Z^b$, where $X$ and $Z$ are the Pauli operators acting on the qudit; these act on a basis $\{|j\rangle,~j=0,1,2,\dots,d-1\}$ for the qudit according to
\begin{eqnarray}
X:|j\rangle&\to& |j+1~({\rm mod}~ d)\rangle~,\nonumber\\
Z:|j\rangle&\to& w^j|j\rangle~,
\end{eqnarray}
where $\omega=\exp(2\pi i/d)$. 

Shifts in $p$ or $q$ can be corrected successfully provided that they satisfy
\begin{equation}
|\Delta q| < \delta/2= \sqrt{\pi\hbar\over 2d}~, \quad |\Delta p| < \delta/2=\sqrt{\pi\hbar\over 2d}~.
\end{equation}
If the shifts in $q$ and $p$ are Gaussian random variables with variance $\sigma^2$, then the probability that a shift causes an uncorrectable error is no larger than the probability that the shift exceeds $\sqrt{\pi\hbar/2d}$, or
\begin{eqnarray}
p_X,p_Z \le & &2\cdot {1\over\sqrt{2\pi\sigma^2}}\int_{\sqrt{\pi\hbar/2d}}^{\infty}dx e^{-x^2/2\sigma^2}
\nonumber\\ &=&{\rm erfc}(\sqrt{\pi\hbar/4d\sigma^2})~,
\label{prob_bound}
\end{eqnarray}
where ${\rm erfc}$ denotes the complementary error function. Here $p_X$ is the probability of an ``$X$ error'' acting on the qudit, of the form $X^a$ for $a\not\equiv 0~({\rm mod}~d)$, and $p_Z$ denotes the probability of a ``$Z$ error'' of the form $Z^b$ for $b\not\equiv 0~({\rm mod}~d)$. The $X$ and $Z$ errors are uncorrelated, and errors with $a,b=\pm 1$ are much more likely than errors with $|a|,|b|>1$. By choosing $d\sim \hbar/\sigma^{2}$, we can achieve a small error probability for each oscillator.

The second stage of the argument is to determine the rate that can be achieved by a qudit code if  $p_X,p_Z$ satisfy eq.~(\ref{prob_bound}). We will consider codes of the Calderbank-Shor-Steane (CSS) type, for which the correction of $X$ errors and $Z$ errors can be considered separately \cite{cal_shor,steane_code}. A CSS code is a stabilizer code, in which each stabilizer generator is either a tensor product of $I$'s and powers of $Z$ (measuring these generators diagnoses the $X$ errors) or a tensor product of $I$'s and powers of $X$ (for diagnosing the $Z$ errors). 

We can establish an achievable rate by averaging the error probability over CSS codes; we give only an informal sketch of the argument. Suppose that we fix the block size $N$ and the number of encoded qudits $k$. Now select the generators of the code's stabilizer group at random. 
About half of the $N-k$ generators are of the $Z$ type and about half are of the $X$ type. Thus the number of possible values for the eigenvalues of the generators of each type is about
\begin{equation}
d^{{1\over 2}(N-k)}~.
\end{equation}
Now we can analyze the probability that an uncorrectable $X$ error afflicts the encoded quantum state (the probability of an uncorrectable $Z$ error is analyzed in exactly the same way). Suppose that $X$ errors act independently on the $N$ qudits in the block, with a probability of error per qudit of $p_X$. Thus for large $N$, the typical number of damaged qudits is close to $p_X\cdot N$. A damaged qudit can be damaged in any of $d-1$ different ways ($X^a$, where $a=1,2, \dots, (d-1)$). We will suppose, pessimistically, that all $d-1$ shifts of the qudit are equally likely. The actual situation that arises in our concatenated coding scheme is more favorable -- small values of $|a|$ are more likely -- but our argument will not exploit this feature.

Thus, with high probability, the error that afflicts the block will belong to a typical set of errors that contains a number of elements close to 
\begin{equation}
N_{\rm typ} \sim {N\choose {Np_X}}(d-1)^{N p_X }\sim d^{N\left(H_d(p_X)+ p_X\log_d(d-1)\right)}~,
\end{equation}
where 
\begin{equation}
H_d(p)= -p\log_d p - (1-p)\log_d(1-p)~.
\end{equation}
If a particular typical error occurs, then recovery will succeed as long as there is no other typical error that generates the same error syndrome. It will be highly unlikely that another typical error has the same syndrome as the actual error, provided that the number of possible error syndromes $d^{{1\over 2}(N-k)}$ is large compared to the number of typical errors. 
Therefore, the $X$ errors can be corrected with high probability for
\begin{eqnarray}
&& {1\over 2}\left(1-{k\over N}\right) \nonumber\\
&& \quad > {1\over N}\cdot \log_d N_{\rm typ}\sim H_d(p_X)+ p_X\log_d(d-1)~,
\end{eqnarray}
or for a rate $R_d$ in qudits satisfying
\begin{equation}
\label{css_rate_x}
R_d\equiv {k\over N} < 1- 2H_d(p_X) - 2p_X\log_d(d-1)
\end{equation}
Similarly, the $Z$ errors can be corrected with high probability by a random CSS code if the rate satisfies 
\begin{equation}
R_d < 1- 2H_d(p_Z) - 2p_Z\log_d(d-1)~.
\label{css_rate_z}
\end{equation}
Converted to qubits, the rate becomes
\begin{equation}
R= \log_2 d \cdot R_d
\label{dits_to_bits}
\end{equation}
Under these conditions, the probability of error averaged over CSS codes becomes arbitrarily small for $N$ large. It follows that there is a particular sequence of CSS codes with rate approaching eq.~(\ref{css_rate_x}-\ref{dits_to_bits}), and error probability going to zero in the limit $N\to\infty$.

\begin{figure}
\begin{center}
\leavevmode
\epsfxsize=3in
\epsfbox{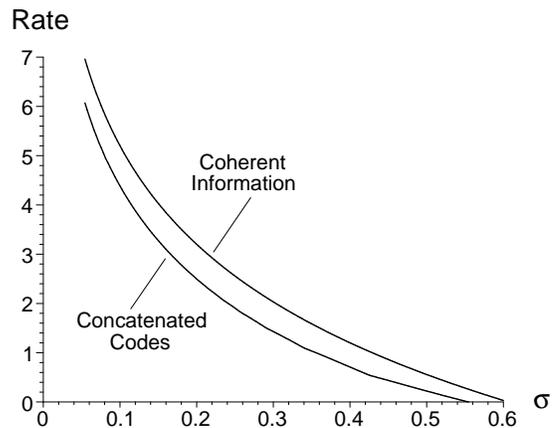}
\end{center}
\caption{Rates achieved by concatenated codes, compared to the one-shot coherent information optimized over Gaussian input states. Here $\sigma$ is the standard deviation of the magnitude of the phase-space displacement introduced by the channel, in units with $\hbar=1$.}
\label{fig:rates}
\end{figure}

\begin{figure}
\begin{center}
\leavevmode
\epsfxsize=3in
\epsfbox{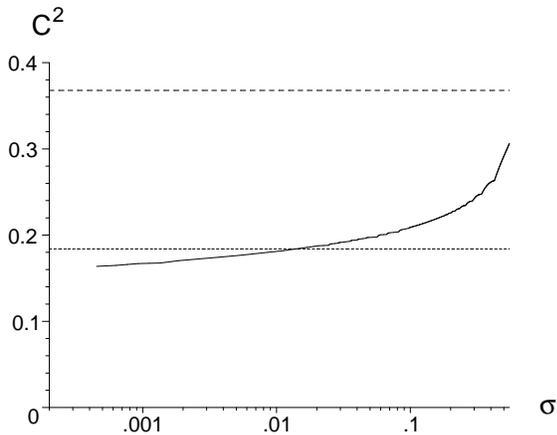}
\end{center}
\caption{The slowly varying function $C^2$, defined by $R=\log_2(C^2/\sigma^2)$, where $R$ is the rate achievable with concatenated codes. Units have been chosen such that $\hbar=1$. The horizontal lines are at $C^2=1/e$, corresponding to a rate equal to the coherent information, and at $C^2=1/2e$, corresponding to one qubit below the coherent information.}
\label{fig:C2}
\end{figure}

For given $\sigma^2$, the optimal rate that can be attained by concatenating a code that encodes a qudit in a single oscillator with a random CSS code, is found by estimating $p_X$ and $p_Z$ using eq.~(\ref{prob_bound}) and then choosing $d$ to maximize the rate $R$ given by eq.~(\ref{css_rate_x}-\ref{dits_to_bits}). The results are shown in Fig.~\ref{fig:rates}. This rate (in qubits) can be expressed as
\begin{equation}
R= \log_2\left(C^2\hbar /\sigma^2\right)~,
\end{equation}
where $C^2$ is a slowly varying function of $\sigma^2/\hbar$ plotted in Fig.~\ref{fig:C2}. It turns out that this rate is actually fairly close to $\log_2 d$; that is, the optimal dimension $d$ of the qudit encoded in each oscillator is approximately $C^2\hbar/\sigma^2$. With this choice for $d$, the error rate for each oscillator is reasonably small, and the random CSS code reduces the error probability for the encoded state to a value exponentially small in $N$ at a modest cost in rate. The rate achieved by concatenating coding lies strictly below the coherent information $I_Q$, but comes within one qubit of $I_Q$ for $\sigma^2 > 1.88 \times 10^{-4}$.

Both the concatenated codes and the codes derived from efficient sphere packings are stabilizer codes, and therefore both are associated with lattices in $2N$-dimensional phase space. But while the sphere-packing codes have been chosen so that the shortest nonzero vector on the lattice is large relative to the size of the unit cell, the concatenated codes correspond to sphere packings of poor quality. For the concatenated codes, the shortest vector of the normalizer lattice has length $\ell$, where  
\begin{equation}
\ell^2=2\pi\hbar/d
\end{equation}
and the rate $R$ is close to $\log_2 d$. The efficient sphere packings have radius $r=\ell/2$ close to $\sqrt{n\sigma^2}$, or
\begin{equation}
\ell^2={8N\hbar\over e} \cdot 2^{-R}~.
\end{equation} 
Hence, if we compare sphere-packing codes and concatenated codes with comparable rates, the 
sphere-packing codes have minimum distance that is larger by a factor of about $\sqrt{4N/\pi e}$.
The concatenated codes achieve a high rate not because the minimum distance of the lattice is large, but rather because the decoding procedure exploits the hierarchical structure of the code.

\section{The classical Gaussian channel}
We have found that quantum stabilizer codes based on efficient sphere packings can achieve rates for the Gaussian quantum channel that match the one-shot coherent information, and that concatenated codes achieve rates that are below, but close to, the coherent information. Now, as an aside, we will discuss the corresponding statements for the classical Gaussian channel. We will see, in particular, that concatenated codes achieve rates that are close to the classical channel capacity.

Shannon's expression for the capacity of the classical Gaussian channel can be understood heuristically as follows \cite{shannon,cover}. If the input signals have average power $P$, which is inflated by the Gaussian noise to $P+\sigma^2$, then if $n$ real variables are transmitted, the total volume occupied by the space of output signals is the volume of a sphere of radius $\sqrt{n(P+\sigma^2)}$, or
\begin{equation}
\label{total_volume}
{\rm total ~volume}= V_n\cdot \left(n(P+\sigma^2)\right)^{n/2}~.
\end{equation}
We will decode a received message as the signal state that is the minimal distance away.
Consider averaging over all codes that satisfy the power constraint and have $m$ signals. When a message is received, the signal that was sent will typically occupy a decoding sphere of radius $\sqrt{(n(\sigma^2 +\varepsilon)}$ centered at the received message, which has volume
\begin{equation}
\label{decoding_volume}
{\rm decoding ~ sphere ~volume}= V_n\cdot \left(n(\sigma^2+\varepsilon)\right)^{n/2}~.
\end{equation}
A decoding error can arise if another one of the $m$ signals, aside from the one that was sent, is also contained in the decoding sphere. The probability that a randomly selected signal inside the sphere of radius $\sqrt{n(P+\sigma^2)}$ is contained in a particular decoding sphere of radius $\sqrt{n(\sigma^2+\varepsilon)}$ is the ratio of the volume of the spheres, so the probability of a decoding error can be upper bounded by $m$ times that ratio, or 
\begin{equation}
P_{\rm error} < m \cdot \left({\sigma^2+\varepsilon\over \sigma^2+P}\right)^{n/2}=\left(2^{2R}\cdot{\sigma^2+\varepsilon\over \sigma^2+P}\right)^{n/2}~,
\end{equation}
where $R$ is the rate of the code. If the probability of error averaged over codes and signals satisfies this bound, there is a particular code that satisfies the bound when we average only over signals. If $P_{\rm error}<\delta$ when we average over signals, then we can discard at most half of all the signals (reducing the rate by at most $1/n$ bits) to obtain a new code with $P_{\rm error}<2\delta$ for {\em all} signals.  Since $\varepsilon$ can be chosen arbitrarily small for sufficiently large $n$, we conclude that there exist codes with arbitrarily small probability of error and rate $R$  arbitrarily close to 
\begin{equation}
C={1\over 2} \log_2\left(1+{P\over\sigma^2}\right)~,
\end{equation}
which is the Shannon capacity. Conversely, for any rate exceeding $C$, the decoding spheres inevitably have nonnegligible overlaps, and the error rate cannot be arbitrarily small.

Suppose that, instead of Shannon's random coding, we use a lattice code based on an efficient packing of spheres. In this case, the power constraint can be imposed by including as signals all lattice sites that are contained in an $n$-dimensional ball of radius $\sqrt{nP}$, and the typical shifts by distance $\sqrt{n\sigma^2}$ must be correctable. Thus decoding spheres of radius $\sqrt{n\sigma^2}$ are to be packed into a sphere of total radius $\sqrt{n(P+\sigma^2)}$. Suppose that the lattice is chosen so that nonoverlapping spheres centered at the lattice sites fill a fraction at least $2^{-(n-1)}$ of the total volume; the existence of such a lattice is established by Minkowski's estimate \cite{minkowski}. Then the number $m$ of signals satisfies 
\begin{equation}
m\cdot V_n \cdot (n\sigma^2)^{n/2} \ge 2^{-(n-1)}\cdot V_n \cdot \left(n(P+\sigma^2)\right)^{n/2}~,
\end{equation}
or
\begin{equation}
m \ge 2^{-n}\left(1+{P\over\sigma^2}\right)^{n/2}~,
\end{equation}
corresponding to the rate
\begin{equation}
R \equiv {1\over n}\cdot \log_2 m={1\over 2}\log_2\left(1+{P\over\sigma^2}\right)-1~,
\end{equation}
which is one bit less than the Shannon capacity. 

Much as in the discussion of quantum lattice codes in Sec.~\ref{sec:improve}, an improved estimate of the achievable rate is obtained if we allow the decoding spheres to overlap \cite{debuda,shell1,shell2,shell3}. In fact, there are classical lattice codes with rate arbitrarily close to the capacity, such that the probability of error, {\em averaged} over signals, is arbitrarily small \cite{shell3}. Unfortunately, though, because of the power constraint, the error probability depends on which signal is sent, and the trick of deleting the worst half of the signals would destroy the structure of the lattice.  Alternatively, it can be shown that for any rate
\begin{equation}
R < {1\over 2}\log_2(P/\sigma^2)~,
\end{equation}
there are lattice codes with maximal probability of error that is arbitrarily small \cite{debuda}. This achievable rate approaches the capacity for large $P/\sigma^2$.

Now consider the rates that can be achieved for the Gaussian classical channel with concatenated coding. A $d$-state system (dit) is encoded in each of $n$ real variables. If each real variable  takes one of $d$ possible values, with spacing $2\Delta x$ between the signals, then a shift by $\Delta x$ can be corrected. By replacing the sum over $d$ values by an integral, which can be justified for large $d$, we find an average power per signal
\begin{equation}
P\sim {1\over 2d\Delta x}\int_{-d\Delta x}^{d\Delta x} x^2 dx= {1\over 3} (d\Delta x)^2~;
\end{equation}
thus the largest correctable shift can be expressed in terms of the average power as
\begin{equation}
\Delta x =\sqrt{3P}/d~.
\end{equation}
For the Gaussian channel with mean zero and variance $\sigma^2$, the probability $p$ of an error in each real variable transmitted is no larger than the probability of a shift by a distance exceeding $\Delta x$, or
\begin{equation}
\label{classical_prob_bound}
p \le {\rm erfc}\left(\sqrt{3P/2d^2\sigma^2}\right)~,
\end{equation}
where ${\rm erfc}$ denotes the complementary error function.

\begin{figure}
\begin{center}
\leavevmode
\epsfxsize=3in
\epsfbox{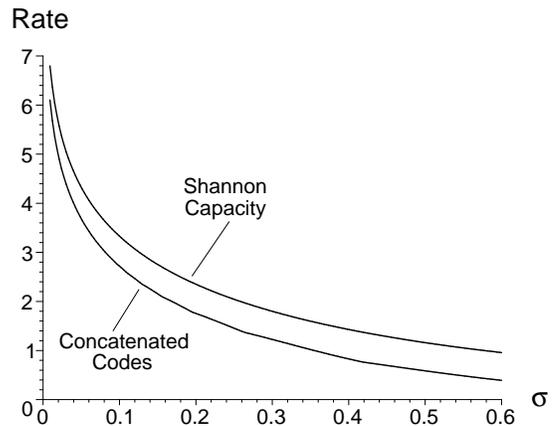}
\end{center}
\caption{Rates for the Gaussian classical channel achievable with concatenated codes, compared to the Shannon capacity. Here $\sigma$ is the standard deviation of the displacement, in units with the power $P=1$.}
\label{fig:classical}
\end{figure}

We reduce the error probability further by encoding $k<n$ dits in the block of $n$ dits. Arguing as in Sec.~\ref{sec:concat}, we see that a random code for dits achieves an asymptotic rate in bits given by 
\begin{equation}
\label{shannon_qudit_rate}
R=\log_2 d \cdot \left( 1- H_d(p) - p \log_d(d-1)\right)~.
\end{equation}
Given $\sigma^2$, using the expression eq.~(\ref{classical_prob_bound}) for $p$, and choosing $d$ to optimize the rate in eq.~(\ref{shannon_qudit_rate}), we obtain a rate close to the Shannon capacity, as shown in Fig.~\ref{fig:classical}. As for the concatenated quantum code, the rate of the concatenated classical code is close to $\log_2 d$, where $d\sim C(\sigma^2)\cdot \sqrt{P/\sigma^2}$, and $C(\sigma^2)$ is a slowly varying function.

\section{Conclusions}

We have described quantum stabilizer codes, based on symplectically integral lattices in phase space, that protect quantum information carried by systems described by continuous quantum variables. With these codes, we can establish lower bounds on the capacities of continuous-variable quantum channels.

For the Gaussian quantum channel, the best rate we know how to achieve with stabilizer coding matches the one-shot coherent information optimized over Gaussian inputs, at least when the value of the coherent information is $\log_2$ of an integer. That our achievable rate matches the coherent information only for isolated values of the noise variance $\sigma^2$ seems to be an artifact of our method of analysis, rather than indicative of any intrinsic property of the channel. Hence it is tempting to speculate that this optimal one-shot coherent information actually is the quantum capacity of the channel. 

Conceivably, better rates can be achieved with {\em nonadditive} quantum codes that cannot be described in terms of symplectically integral lattices. We don't know much about how to construct these codes, or about their properties.

In the case of the depolarizing channel acting on qubits, Shor and Smolin discovered that rates exceeding the one-shot coherent information could be achieved. Their construction used concatenated codes, where the ``outer code'' is a random stabilizer code, and the ``inner code'' is a degenerate code with a small block size \cite{shor_smolin}. The analogous procedure for the Gaussian channel would be to concatenate an outer code based on a symplectically integral lattice with an inner code that encodes one logical oscillator in a block of several oscillators. This inner code, then,  embeds an infinite-dimensional code space in a larger infinite-dimensional space, as do codes constructed by Braunstein \cite{braunstein} and Lloyd and Slotine \cite{lloyd_slot}. However, we have not been able to find concatenated codes of this type that achieve rates exceeding the one-shot coherent information of the Gaussian channel.

\acknowledgments

We thank Dave Beckman, Anne-Marie Berg\'e, Bob McEliece, Michael Postol, Eric Rains, Peter Shor, and Edward Witten for helpful discussions and correspondence. This work has been supported in part by the Department of Energy under Grant No. DE-FG03-92-ER40701, by the National Science Foundation under Grant No. EIA-0086038, by the Caltech MURI Center for Quantum Networks under ARO Grant No. DAAD19-00-1-0374, and by an IBM Faculty Partnership Award.

\end{document}